# Vortices and chirality of magnetostatic modes in quasi-2D ferrite disk particles

E.O. Kamenetskii

Ben-Gurion University of the Negev, Beer Sheva 84105, Israel

February 14, 2007

**Abstract**

In this paper we show that the vortex states can be created not only in magnetically soft "small" (with the dipolar and exchange energy competition) cylindrical dots, but also in magnetically saturated "big" (when the exchange is neglected) cylindrical dots. A property associated with a vortex structure becomes evident from an analysis of confinement phenomena of magnetic oscillations in a ferrite disk with a dominating role of magnetic-dipolar (non-exchange-interaction) spectra. In this case the scalar (magnetostatic-potential) wave functions may have a phase singularity in a center of a dot. A non-zero azimuth component of the flow velocity demonstrates the vortex structure. The vortices are guaranteed by the chiral edge states of magnetic-dipolar modes in a quasi-2D ferrite disk.

**1. Introduction**

Magnetically soft ferromagnetic materials generally form domain structures to reduce their magnetostatic (MS) energy. In this context, closure domains are especially suitable. Such magnetic objects are characterized by a closed flux circuit having no magnetic flux leakage outside the material. In very small systems, however, the formation of domain walls is not energetically favored. Specifically, in a dot of a ferromagnetic material of micrometer or submicrometer size, a curling spin configuration – that is, a magnetization vortex – has been proposed to occur in place of domains. The vortex consists of an in-plane, flux-closure magnetization distribution and a central core whose magnetization is perpendicular to the dot plane. It has been shown that under certain conditions a vortex structure will be stable because of competition between the exchange and dipole interactions. A magnetic vortex means the "curling magnetization distribution". One obtains the clockwise and counter-clockwise rotations of magnetization vector $\vec{m}$ in the dot plane [1 – 3].

In a general case, both short-range exchange and long-range dipole-dipole (magnetostatic) interactions contribute to eigenfrequencies of the collective spin excitation. The importance of the MS energy increases gradually as the particle size increases. MS ferromagnetism has a character essentially different from exchange ferromagnetism [4, 5]. This statement finds strong confirmation in confinement phenomena of magnetic-dipolar-mode (MDM) oscillations. The dipole interaction provides us with a long-range mechanism of interaction, where a magnetic medium is considered as a continuum. Contrary to an exchange spin wave, in magnetic-dipolar waves the local fluctuation of magnetization does not propagate due to interaction between the neighboring spins. When field differences across the sample become comparable to the bulk demagnetizing fields the local-oscillator approximation is no longer valid, and indeed under certain circumstances, entirely new spin dynamics behavior can be observed. This dynamics behavior is the following. Precession of magnetization about a vector of a bias magnetic field produces a small oscillating magnetization $\vec{m}$ and a resulting dynamic demagnetizing field $\vec{H}$, which reacts back on the precession, raising the resonant frequency. In the continuum approximation, vectors $\vec{H}$ and $\vec{m}$ are coupled by the differential relation:

$$\vec{\nabla} \cdot \vec{H} = -4\pi \vec{\nabla} \cdot \vec{m}, \qquad (1)$$

where

$$\vec{H} = -\nabla \psi \qquad (2)$$

and $\psi$ is a MS potential. This, together with the Landau-Lifshitz (LL) equation, leads to complicated integro-differential equations for the mode solutions in a lossless magnetic sample.

For calculation, the formulation based on the MS-Green-function integral problem for magnetization $\vec{m}$ was suggested and used in [6 – 8]. In this case one solves "pure static" MS equations for the dipolar field. It is supposed that the sources of a MS field are both volume and surface "magnetic charges" arising from $\nabla \cdot \vec{m}$ and from discontinuity of the normal component of $\vec{m}$ on the surface of a ferrite sample. The MS potential $\psi$ is defined based on integration of the MS Poisson equation

$$\nabla^2 \psi = -4\pi \rho_m, \qquad (3)$$

where

$$\rho_m \equiv -\vec{\nabla} \cdot \vec{m}. \qquad (4)$$

For the edge and volume sources one obtains [9]:

$$\psi = -\int_V \frac{\vec{\nabla}_{r'} \cdot \vec{m}(\vec{r}\,')}{|\vec{r} - \vec{r}\,'|} dV + \oint_S \frac{\vec{n}\,' \cdot \vec{m}(\vec{r}\,')}{|\vec{r} - \vec{r}\,'|} dS. \qquad (5)$$

Such a theoretical analysis of the RF magnetization eigenvalue problem encounters, however, a significant difficulty due to the absence of exact information of the boundary conditions for RF magnetization $\vec{m}$. It is well known that in classical electrodynamics the boundary conditions are imposed on the normal components of the magnetic induction and the tangential component of the magnetic fields, but not on the components of magnetization. So the dynamic magnetization at the boundary of a magnetic element is undefined from classical electrodynamics. To derive dipolar boundary conditions for dynamic magnetization, one calculates the MS energy arising from the effective magnetic charge at the magnetic element. The obtained phenomenological pinning parameter for RF magnetization results from the induced surface "magnetic charges" at the edges of a finite-size non-elipsoudal magnetic element. The physical meaning of this pinning parameter is to minimize the surface MS energy. To solve the eigenvalue problem for magnetic-dipolar magnetization one considers the inhomogeneous dipolar field obtained directly from magnetostatic equation (5) and uses the vector torque equation for magnetization [6 – 8]. The solution of the problem yields a discrete set of eigenfunctions. It has been concluded that these eigenfunctions of magnetization are orthogonal with real eigenvalues because the kernel of the Fredholm-type integral equation is symmetric and real.

An analysis gives a good correlation with the experimental results for "small" (submicron size) magnetic particles. At the same time, the approach cannot be considered generally as well suited to describe spectral characteristics of MDMs in a "big" magnetic particle. It is known that in a case of spectral problems, for integral equation with positive definite symmetrical kernel there should be corresponding Hermitian differential operator (see e.g. [10]). Now the question arises: What are the Hermitian-operator differential equations corresponding to the integral MS operators in Refs. [6 – 8]? Certainly, the Poisson equation (3) is not a Hermitian-operator differential equation. Also there are no corresponding Hermitian-operator differential equations written for the RF MDM (non-exchange-interaction) magnetization $\vec{m}$. One cannot use the short-range exchange differential operator to solve the far-range dipolar boundary problem. So in the MDM problem one cannot expand the RF magnetization by the eigenfunctions of the exchange operator. The MS spectral problem, being analyzed based on the above approach, is not the self-conjugate problem and statements about completeness of the magnetization eigenfunctions [6 – 8] may cast certain doubts. The dipolar-contribution factor becomes also a very complicated problem for calculation of the magnetic normal modes based on micromagnetic simulations. In this case the dipolar energy is considered as the most difficult part to treat, compared to other types of the magnetic energy. The Poisson-equation approach is not well suited to yield the derivatives needed for the dynamical magnetization torque matrix [11].

The flaw of the above integral eigenvalue problem becomes evident for "big" (with sizes tens of micrometers and more) ferrite samples. In the continuum approximation, to get correct solutions for dynamical processes inside a "big" ferrite particle one has to presuppose existence of certain retardation effects for the MS fields. For "big" magnetic samples, an analysis of the egenvalue problem based on the differential-operator equation for a fictitious MS-potential *wave* function $\psi$ may eliminate the above difficulties of the MS-Green-function integral problem. In this case one uses the continuum approximation based on the known [from the linearized local (non-exchange-interaction) LL equation] permeability tensor $\vec{\mu}$ [12]. One supposes that there is a spectral problem for the MS-potential propagating fields which cause and govern propagation of magnetization fluctuations. In other words, space-time magnetization fluctuations are corollaries of the propagating MS-potential fields, but there is no the magnetization-wave spectral problem. The boundary conditions are imposed on the MS-potential field and not on the RF magnetization [13]. Usually, to calculate these effects the Walker's [12] differential formulation is used and the general solution of this equation is expressed through the MS-potential *wave* function $\psi$:

$$\vec{\nabla} \cdot (\vec{\mu} \cdot \vec{\nabla} \psi) = 0. \qquad (6)$$

Eq. (6) follows immediately from Eqs. (1) and (2). The way of solution based on the Walker's equation is used both for continuous-wave FMR [12] and NMR [14] measurements. The boundary conditions one obtains immediately from an analysis of the spectral problem for MS-potential wave function $\psi$.

In this paper we show that the vortex states can be created not only in magnetically soft "small" (with the dipolar and exchange energy competition) cylindrical dots, but also in magnetically saturated "big" (when the exchange is neglected) cylindrical dots. Our definition of the long-range order is based on the overall properties of the system rather than on the behavior of a two-point correlation function. A property associated with a vortex structure becomes evident from an analysis of the spectral problem for MS-potential wave function $\psi$. We show that in a case of a normally magnetized thin-film ferrite disk, MDM vortices can be guaranteed by chirality of the MS waves. The dynamical symmetry breaking in MS-wave oscillations shows special gauge transformation of the fields.

In spite of the fact that vortices can appear in different kinds of physical phenomena, yet such "swirling" entities seem to elude an all-inclusive definition. It appears that the characters of magnetic vortices in "big" saturated disks and in "small" soft disks are very different. A magnetization vortex in a magnetically soft sample cannot be characterized by some invariant, such as the flux of vorticity. So a vorticity thread may not be defined for the magnetization vortex [15]. At the same time, the MDM vortices take place due to a non-zero azimuth component of the flow velocity. It is well-known that in quantum systems, the gradient of the scalar potential plays the role analogous to the velocity. The circulation of the velocity therefore becomes quantized. Our analysis demonstrates such quantum-like vortices in normally magnetized MDM ferrite disks.

**2. Energy relations for MDMs in a normally magnetized ferrite disk**

It was supposed that expression for power flow density for MS waves can be derived with formal use of the Maxwell equations for the curl electric field and the potential magnetic field [16]. This may give, in particular, a foundation for well-known theories of the MS-wave excitation by an electric current (see e.g. [17, 18]). Based on this expression, Boardman *et al* [19] considered the power-flow-density rotation excited by three planar antennas to analyze possible vortex structures of MS waves in a ferrite film. It looks, however, that this way of an analysis may be used as a proper approximation for the main MS mode but reveals a clear physical contradiction for a general MS-wave spectral problem. Concerning the MS-wave propagation effects, it was disputed in [20, 21] that from a classical electrodynamics point of view one does not have a physical mechanism describing the effect of transformation of the curl electric field to the potential magnetic field. Also the gauge transformation in this derivation does not fall under the known gauge transformations, neither the Lorentz gauge nor the Coulomb gauge, and cannot formally lead to the wave equation [21].

MS waves can propagate only due to ferrite-medium confinement phenomena [12, 16]. So the power flow density of MS waves should be considered via an analysis of the mode spectral problem in a certain waveguide structure, but not based on an analysis of the wave propagation in a boundless magnetic medium. It means that the MS-wave power flow density should have the only physical meaning as a norm of a certain propagating mode in a magnetic waveguide structure. As a necessary consequence, this leads to the question of orthogonality and completeness of MS modes in such a waveguide. In this problem MS-potential $\psi$ acquires a special physical meaning as a scalar wave function in a Hilbert functional space. Such a standpoint is especially important for study of the MS-wave vortices: in confined magnetic structures, vortices should be characterized as stable and energy favored "swirling" entities.

For MS waves one can use the following operator equation [22]

$$\hat{L} V = 0, \qquad (7)$$

where

$$\hat{L} \equiv \begin{pmatrix} (\ddot{\mu})^{-1} & \nabla \\ -\nabla \cdot & 0 \end{pmatrix} \qquad (8)$$

is the differential-matrix operator, $\ddot{\mu}$ is the permeability tensor, and

$$V \equiv \begin{pmatrix} \vec{B} \\ \psi \end{pmatrix} \qquad (9)$$

is the vector function included in the domain of definition of operator $\hat{L}$. Outside of a ferrite region one has the same equations but with $\vec{\vec{\mu}} = \vec{\vec{I}}$, where $\vec{\vec{I}}$ is the unit matrix.

For MS-wave propagation in an infinite ferrite rod along $z$ axis one has

$$(\hat{L}_\perp - i\beta_z \hat{R})\tilde{V} = 0, \qquad (10)$$

where

$$\hat{L}_\perp \equiv \begin{pmatrix} (\vec{\vec{\mu}})^{-1} & \nabla_\perp \\ -\nabla_\perp \cdot & 0 \end{pmatrix}, \qquad (11)$$

subscript $\perp$ means differentiation over a waveguide cross section, $\beta_z$ is the MS-wave propagation constant along $z$ axis,

$$\tilde{V} \equiv \begin{pmatrix} \tilde{\vec{B}} \\ \tilde{\psi} \end{pmatrix}, \qquad (12)$$

is the membrane vector function ($V \equiv \tilde{V} e^{-i\beta_z z}$),

$$\hat{R} \equiv \begin{pmatrix} 0 & \vec{e}_z \\ -\vec{e}_z & 0 \end{pmatrix}, \qquad (13)$$

$\vec{e}_z$ is a unit vector along the axis of the wave propagation.

Integration by parts on $S$ – a square of an open MS-wave cylindrical waveguide – of the integral $\int_S (\hat{L}_\perp \tilde{V})\tilde{V}^* dS$ gives the contour integral in a form $\oint_C (\tilde{B}_r \tilde{\psi}^* - \tilde{B}_r^* \tilde{\psi})\, dC$, where $C$ is a contour surrounding a cylindrical ferrite core and $\tilde{B}_r$ is a radial component of a membrane function of the magnetic flux density. Operator $\hat{L}_\perp$ becomes self-adjoint for homogeneous boundary conditions (continuity of $\tilde{\psi}$ and $\tilde{B}_r$) on contour $C$. Based on the homogeneous boundary conditions one obtains the orthogonality relation for MDMs:

$$(\beta_p - \beta_q)\int_S (\hat{R}\tilde{V}_p)(\tilde{V}_q)^* dS = 0. \qquad (14)$$

The norm of mode $p$ is determined as

$$N_p = \int_S \left( \tilde{\psi}_p \tilde{\vec{B}}_p^* - \tilde{\psi}_p^* \tilde{\vec{B}}_p \right) \cdot \vec{e}_z dS. \qquad (15)$$

It is easy to show that norm $N_p$, being multiplied by a proper dimensional coefficient, corresponds to the power flow of the waveguide mode $p$ through a waveguide cross section. For monochromatic fields with time variation $\sim e^{i\omega t}$ we have for the power flow in Gaussian units:

$$P_p = -\frac{i\omega}{16\pi} N_p = \frac{i\omega}{16\pi} \int_S \left( \tilde{\psi}_p^* \tilde{\vec{B}}_p - \tilde{\psi}_p \tilde{\vec{B}}_p^* \right) \cdot \vec{e}_z dS . \tag{16}$$

The proof of this fact is evident. Based on Eq. (7) [together with the equation complex conjugated with Eq. (7)] and taking into account expression $\vec{B} = -\vec{\mu}(\omega)\nabla\psi$, one obtains

$$-\frac{i\omega}{16\pi} \nabla \cdot \left( \psi \vec{B}^* - \psi^* \vec{B} \right) = -\frac{i\omega}{16\pi} \left[ \vec{B} \cdot \left( \vec{\mu}^*(\omega) \right)^{-1} \cdot \vec{B}^* - \vec{B}^* \cdot \left( \vec{\mu}(\omega) \right)^{-1} \cdot \vec{B} \right]. \tag{17}$$

This is an energy balance equation for monochromatic MS waves in lossy magnetic media. In the right-hand side of this equation we have the average density of magnetic losses taken with an opposite sign. Thus the term in the left-hand side is the divergence of the power flow density. Really, in a region of a FMR, the average density of magnetic energy absorption is expressed as [23]

$$\left\langle \frac{\partial w}{\partial t} \right\rangle_{abs} = \frac{i\omega}{2} \vec{H}^* \cdot \vec{\chi}^{ah}(\omega) \cdot \vec{H} , \tag{18}$$

where $\vec{H}$ is an internal RF magnetic field, $\vec{\chi}$ is the magnetic susceptibility tensor, superscript *ah* means "anti-Hermitian". This expression can be rewritten as

$$\left\langle \frac{\partial w}{\partial t} \right\rangle_{abs} = \frac{i\omega}{8\pi} \vec{H}^* \cdot \vec{\mu}^{ah}(\omega) \cdot \vec{H} . \tag{19}$$

It can be easily shown that

$$\vec{B} \cdot \left( \vec{\mu}^*(\omega) \right)^{-1} \cdot \vec{B}^* - \vec{B}^* \cdot \left( \vec{\mu}(\omega) \right)^{-1} \cdot \vec{B} = 2\vec{H}^* \cdot \vec{\mu}^{ah}(\omega) \cdot \vec{H} . \tag{20}$$

So the right-hand side of Eq. (17) describes the density of magnetic losses taken with an opposite sign.

We introduced magnetic losses in the above analysis to be able to define clearly the power flow density of the MS-wave waveguide mode. The next question is about the average density of accumulated energy for MS-wave modes. For this problem we will analyze a lossless structure. For a bias magnetic field directed along *z* axis one has the permeability tensor written in a form:

$$\vec{\mu} = \begin{pmatrix} \mu & i\mu_a & 0 \\ -i\mu_a & \mu & 0 \\ 0 & 0 & 1 \end{pmatrix}, \tag{21}$$

where $\mu$ and $\mu_a$ are real quantities [12]. Components of permeability tensor $\tilde{\tilde{\mu}}$ are dependent on both frequency $\omega$ and bias magnetic field $H_0$. This may lead to unique features in spectral problems of MS oscillations. For given mode $p$ in an axially magnetized ferrite rod and taking into account Eq. (21), we rewrite now Eq. (6) as follows

$$\left(\hat{G}_\perp - (\beta_{z_p})^2\right)\tilde{\psi}_p = 0, \tag{22}$$

where

$$\hat{G}_\perp \equiv \mu \nabla_\perp^2, \tag{23}$$

$\nabla_\perp^2$ is the two-dimensional (with respect to cross-sectional coordinates) Laplace operator. Operator $\hat{G}_\perp$ is *positive definite* for negative quantities $\mu$. Outside a ferrite region Eq. (22) becomes the Laplace equation ($\mu = 1$). Double integration by parts on square $S$ – a cross-section of a waveguide structure – of the integral $\int_S (\hat{G}_\perp \tilde{\psi})\tilde{\psi}^* dS$ gives the boundary conditions for self-adjointess of operator $\hat{G}_\perp$. For given frequency $\omega$ in a region of a bias magnetic field $H_0$ where $\mu < 0$, one has a complete orthonormal basis of eigen modes $\tilde{\psi}$ of operator $\hat{G}_\perp$. Eigen numbers are quantities $(\beta_{z_p})^2$ [21, 22].

Let us represent membrane function $\tilde{\psi}_p$ as

$$\tilde{\psi}_p = C_p \tilde{\varphi}_p, \tag{24}$$

where $C_p$ is a dimensional normalization coefficient and $\tilde{\varphi}_p$ is a dimensionless membrane function for mode $p$ [24]. We also formally introduce now a certain mode quantity $E_p$ which is defined as

$$E_p \equiv \frac{g_p}{16\pi}(\beta_{z_p})^2. \tag{25}$$

Here $g_p$ is another dimensional normalization coefficient for mode $p$. The physical meaning of quantities $E_p$ and $g$ will be shown and discussed later on. Based on relation (25) and taking into account Eq. (24) one rewrites Eq. (22) in a form:

$$\hat{F}_\perp \tilde{\varphi}_p = E_p \tilde{\varphi}_p, \tag{26}$$

where

$$\hat{F}_\perp \equiv \frac{g_p}{16\pi}\mu \nabla_\perp^2. \tag{27}$$

Assuming self-adjointness of operator $\hat{G}_\perp$ (and, consequently, self-adjointness of operator $\hat{F}_\perp$) one obtains the following orthonormality conditions

$$(E_p - E_q)\int_S \tilde{\varphi}_p \tilde{\varphi}_q^* \, dS = 0. \tag{28}$$

In this connection, we will write

$$\int_S \tilde{\varphi}_p \tilde{\varphi}_q^* \, dS = \delta_{pq}, \tag{29}$$

where $\delta_{pq}$ is the Kronecker delta.

Taking into account Eq. (21), one has the Walker equation for an axially magnetized ferrite rod:

$$\mu \nabla_\perp^2 \psi + \nabla_\parallel \psi = 0, \tag{30}$$

where subscript $\parallel$ means differentiation along z axis. Based on Eq. (30), one rewrites Eq. (26) as

$$\hat{F}_\parallel \tilde{\varphi}_p = -E_p \tilde{\varphi}_p, \tag{31}$$

where

$$\hat{F}_\parallel \equiv \frac{g_p}{16\pi} \nabla_\parallel^2. \tag{32}$$

For given mode $p$, Eq. (31) looks like the time-independent one-dimensional Schrödinger equation for a free particle [25]. For two modes, $p$ and $q$, one obtains:

$$(E_p - E_q) \tilde{\varphi}_p \tilde{\varphi}_q^* = \nabla_\parallel \cdot \left( \tilde{\varphi}_p \nabla_\parallel \tilde{\varphi}_q^* - \tilde{\varphi}_q^* \nabla_\parallel \tilde{\varphi}_p \right). \tag{33}$$

Integration of this equation over cross-section $S$ gives immediately the orthogonality condition (30). Based on formal resemblance with the quantum mechanics description, MS-mode propagation in an infinite ferrite rod can be considered as a "free-particle motion" with "potential energy" constant along $z$ axis.

Now let us consider a ferrite disk normally magnetized along $z$ axis. For a ferrite disk with thickness $d$, one has restrictions of a "potential energy" in $z$ direction. At $z = 0, d$ there are the boundary conditions: continuity of $\tilde{\varphi}$ and $\nabla_\parallel \tilde{\varphi}$ [22, 24]. For given mode $p$, there is a specific quantity of $E_p$ and a specific height of an effective confining potential $U_p^{eff}$. In a case of a ferrite disk, Eq. (31) should be represented as

$$\left( \frac{d^2}{dz^2} + \frac{16\pi}{g_p} E_p \right) \tilde{\varphi}_p = 0 \tag{34}$$

for $0 \leq z \leq d$ and

$$\left[\frac{d^2}{dz^2}+\frac{16\pi}{g_p}\left(E_p-U_p^{eff}\right)\right]\tilde{\varphi}_p=0 \tag{35}$$

for $z \leq 0; z \geq d$. Inside a ferrite disk ($0 \leq z \leq d$) one has

$$(\beta_{z_p})^2 = \frac{16\pi}{g_p}E_p, \tag{36}$$

while outside a disk ($z \leq 0; z \geq d$), for $E_p < U_p^{eff}$, one has [24]

$$\tilde{\varphi}_p = const \cdot e^{\mp \alpha_p z}, \quad \text{where} \quad (\alpha_p)^2 = \frac{16\pi}{g_p}\left(U_p^{eff}-E_p\right). \tag{37}$$

Here signs $\mp$ correspond to regions $z \leq 0$ and $z \geq d$, respectively. It follows (see Ref.[22]) that $\alpha = \frac{1}{\sqrt{-\mu}}\beta$, where $\mu$ is a negative quantity. After some transformations, one obtains from the above expressions

$$\frac{U^{eff}}{E} = 1 + \frac{1}{-\mu}. \tag{38}$$

For a FMR region with negative $\mu$ [12], a character of dependence $\frac{U^{eff}}{E} = \frac{U^{eff}}{E}(H_0)$, where $H_0$ is a bias magnetic field, is shown in Fig. 1. In this figure, the boundary quantities of a bias magnetic field are [12]: $H_{0_1} = \frac{\omega}{\gamma}; H_{0_2} = \left[\left(\frac{\omega}{\gamma}\right)^2 + 2\pi M_0\right]^{1/2} - 2\pi M_0$, where $\gamma$ is the gyromagnetic ratio and $M_0$ is the saturation magnetization. Based on the known positions of MDM resonances in a ferrite disk with respect to $H_0$ [24], one can fit proper mode quantities of $\frac{U^{eff}}{E}$. Bias magnetic fields corresponding to different MDMs (*p, q, r*) are designated in Fig, 1 as $H_0^{(p)}, H_0^{(q)}, H_0^{(r)}$. For $r > p > q$, there are $H_0^{(p)} > H_0^{(q)} > H_0^{(r)}$. Qualitative pictures of the $U^{eff}$ and $E$ levels for three modes *p, q, r* in "potential wells" are shown in Fig. 2. One has $U_p^{eff} < U_q^{eff} < U_r^{eff}$ and $E_p < E_q < E_r$.

Further comparison with quantum mechanics approach leads to understanding real physical meaning of the above formal analysis. From the principle of superposition of states, it follows that wave functions $\tilde{\varphi}_p$ ($p = 1,2,...$), describing our "quantum" system, are "vectors" in an abstract space of an infinite number of dimensions – the Hilbert space. In quantum mechanics, this is the case of so-called *energetic representation*, when the system energy runs through a discrete sequence of values. In the energetic representation, a square of a modulus of the wave function defines probability to find a system with a certain energy value [25, 26]. In our case, scalar-wave membrane function $\tilde{\varphi}$ can be represented as

$$\tilde{\varphi} = \sum_p a_p \tilde{\varphi}_p \qquad (39)$$

and the probability to find a system in a certain state *p* is defined as

$$\left|a_p\right|^2 = \left|\int_S \tilde{\varphi}\, \tilde{\varphi}_p^* dS\right|^2. \qquad (40)$$

The complete-set orthonormalized basic vectors in Hilbert space describe a stationary "ensemble". In an infinite axially magnetized ferrite rod one can consider the case of the evolution of the "ensemble". The MS-potential wave function $\psi$ entirely defines the state of our system – the magnetic sample. It means that representation of this function in a certain time moment not only describes the system behavior at the present moment, but defines the behavior in all future time moments. Mathematically, it means that (taking into account the principle of superposition) there should be [25]:

$$i\frac{\partial \psi}{\partial t} = \hat{Q}\, \psi, \qquad (41)$$

where $\hat{Q}$ is a certain linear operator. From this equation it can be shown that for orthonormalized basic vectors, operator $\hat{Q}$ is a self-conjugate differential operator. Thus Eq. (41) is a wave equation for complex scalar wave function $\psi$ [25].

Let us represent a MS-potential function of mode *p* propagating in an infinite axially magnetized ferrite rod as a quasi-monochromatic quantity:

$$\Psi_p(z,t) = \psi_p^{(\max)}(z,t)\, e^{i(\omega t - \beta_{z_p} z)}, \qquad (42)$$

where "quasi-membrane" complex amplitude $\psi_p^{(\max)}(z,t)$ is a smooth function of the longitudinal coordinate and time, so that

$$\left|\frac{1}{\beta_{z_p}} \frac{\partial \psi_p^{(\max)}}{\partial z}\right| \ll \psi_p^{(\max)}, \qquad \left|\frac{1}{\omega} \frac{\partial \psi_p^{(\max)}}{\partial t}\right| \ll \psi_p^{(\max)}. \qquad (43)$$

Situation of the quasi-monochromatic behavior can be realized, in particular, by means of a time-dependent bias magnetic field slowly varying with respect to the Larmor frequency. In this case, a spin-polarized ensemble will adiabatically follow the bias magnetic field and the resulting energy of interaction with a bias field becomes time dependent.

We represent now the quasi-monochromatic MS field described by Eq. (42) as

$$\Psi_p(z,t) = C_p \Phi_p(z,t), \qquad (44)$$

where $C_p$ is a dimensional normalization coefficient [the same as in Eq. (24)] and $\Phi_p(z,t)$ is a dimensionless MS-potential function. Eq. (41) can be written as:

$$i\frac{\partial \Phi_p}{\partial t} = \frac{\omega}{(\beta_{z_p})^2}\nabla_\parallel^2 \Phi_p. \tag{45}$$

The form of operator $\hat{Q} \equiv \frac{\omega}{(\beta_{z_p})^2}\nabla_\parallel^2$ follows from the "stationary-state" conditions. Taking into account Eq. (25), one sees that Eq. (45) has a physical meaning as a quasi-monochromatic generalization of the "stationary-state" equation (31) describing MS modes propagating in a ferrite rod. The "stationary-states" correspond to monochromatic fields with time variation $\sim e^{i\omega t}$. Based on Eq. (45) and its complex conjugated form one easily obtains:

$$\frac{\partial |\Phi_p|^2}{\partial t} = \frac{i\omega}{(\beta_{z_p})^2}\nabla_\parallel \cdot \left(\Phi_p \nabla_\parallel \Phi_p^* - \Phi_p^* \nabla_\parallel \Phi_p\right). \tag{46}$$

For a quasi-monochromatic field, the energy balance equation for waveguide mode $p$ propagating along $z$ axis in a lossless ferrite rod is written as

$$\int_z dz \int_S \frac{\partial \langle w_p \rangle_{acc}}{\partial t} dS = -\int_z dz \int_S \nabla_\parallel \cdot \vec{p}_p\, dS, \tag{47}$$

where $\langle w_p \rangle_{acc}$ is the average density of accumulated magnetic energy of MS mode $p$ and $\vec{p}_p$ is the power flow density of mode $p$. Based on Eqs. (16) and (21) we have for an infinite ferrite rod

$$\frac{\partial \langle w_p \rangle_{acc}}{\partial t} = \frac{i\omega}{16\pi}\nabla_\parallel \cdot \left(\Psi_p \nabla_\parallel \Psi_p^* - \Psi_p^* \nabla_\parallel \Psi_p\right). \tag{48}$$

Comparison of two equations, Eq. (46) and Eq. (48), with taking into account Eq. (44) gives the following relation:

$$\langle w_p \rangle_{acc} = \frac{C_p^2 (\beta_{z_p})^2}{16\pi}|\Phi_p|^2. \tag{49}$$

One can rewrite this equation as

$$\langle w_p \rangle_{acc} = \frac{C_p^2}{g_p} E_p |\tilde{\varphi}_p|^2. \tag{50}$$

Taking into account Eq. (29), one has

$$\int_S \langle w_p \rangle_{acc} dS = \frac{C_p^2}{g_p}\int_S E_p dS. \tag{51}$$

Condition $g_p \equiv C_p^2$ leads us to important conclusion about a physical meaning of quantities $E_p$ and $g_p$ introduced formally above, in Eq. (25). The quantity $g_p$ should be considered as a dimensional normalization coefficient with the same dimension as squared MS-mode-amplitude coefficient $C_p$. The quantity $E_p$ means an average density of accumulated magnetic energy of MS mode *p* propagating in a ferrite rod.

From quantum mechanics, it is known that based on the one-dimensional time-dependent Schrödinger equation, $i\hbar \frac{\partial \phi}{\partial t} = -\frac{\hbar^2}{2m} \nabla_\parallel^2 \phi + U\phi$, one obtains the continuity equation:

$$\frac{\partial |\phi^2|}{\partial t} = -\nabla_\parallel \cdot \vec{j}, \qquad (52)$$

where

$$\vec{j} = \frac{i\hbar}{2m}(\phi \nabla_\parallel \phi^* - \phi^* \nabla_\parallel \phi) \qquad (53)$$

is the probability flow density. One can see a clear correlation between the scalar-wave-function Eq. (46) and one-dimensional quantum mechanics Eqs. (52), (53), when coefficient $\frac{\omega}{(\beta_{z_p})^2}$ is formally replaced by $-\frac{\hbar}{2m}$. This also corresponds to the de Broglie dispersion relation for a particle of effective mass $m_p$ and moment $\hbar \beta_{z_p}$:

$$\omega = \frac{\hbar}{2m_p}(\beta_{z_p})^2, \qquad (54)$$

where the effective mass $m_p$ is a *negative* quantity. This relation shows that there should be the negative frequency $\omega$. The physical meaning of such "negativeness" follows from the negative-dispersion character of MS modes propagating in an axially magnetized ferrite rod [27].

One may wonder whether the MDMs are quantum or classical. We cannot answer this question in a simple question, but give some remarks on it as follows. In our case we have evident both classical and quantum attributes. In the Maxwell theory, we can define a positive-definite energy density, while cannot define a positive-definite probability density. The above analysis shows that every MS mode propagating in a ferrite rod and every MS mode oscillating in a ferrite disk can be characterized by probability density and are described by the Schrödinger-like equation. Propagation of MS modes can be considered as propagation of quasiparticles – the light magnons (LMs). The meaning of the term "light magnon" arises from the fact that the quantities of effective masses of these quasiparticles are much less than effective masses of the ("real", "heavy") magnons – the quasiparticles existing due to the exchange interaction [28]. In our description of MS oscillations we neglect the exchange interaction and the "magnetic stiffness" is characterized by the "weak" dipole-dipole interaction. Expression (54) looks very similar to an effective mass of the "heavy" magnon for exchange-interaction spin waves with the quadratic character of dispersion [12]. The above analysis shows the way of transition from the classical theory to the quantum-like theory of MDMs in a ferrite sample. When one selects properly normalization of a linear combination of differential-equation solutions for MS-potential functions, one may attain coincidence between MS field energy in a sample and energy of an oscillator system. LMs are long-range elementary magnetic excitations which appear due to dipole-dipole interactions in confined-system magnetically ordered continuum. Every oscillating MDM in a ferrite sample is characterized by a certain type of the LMs. There are different LM effective masses for different oscillating MDMs. So in a finite ferrite sample, LMs related to different MDMs (or, in other words, related to different energy states) are not identical particles [28].

## 3. Self-adjointess of differential operators and boundary conditions

In the above consideration, we supposed self-adjointess of both differential operators, $\hat{G}_\perp$ and $\hat{L}_\perp$. Nevertheless, boundary conditions for self-adjointess of operator $\hat{G}_\perp$ differ from boundary conditions necessary for self-adjointess of operator $\hat{L}_\perp$. The boundary conditions for self-adjointess of operator $\hat{G}_\perp$ one defines based on double integration by parts on square $S$ – a cross-section of a waveguide structure – of the integral $\int_S (\hat{G}_\perp \tilde{\psi}) \tilde{\psi}^* dS$. One can see that for a ferrite cylindrical sample of radius $\Re$, these boundary conditions on a lateral surface presumes continuity of MS wave function $\tilde{\psi}$ as well as the relation for derivatives:

$$\mu \left( \frac{\partial \tilde{\psi}}{\partial r} \right)_{r=\Re^-} - \left( \frac{\partial \tilde{\psi}}{\partial r} \right)_{r=\Re^+} = 0. \tag{55}$$

This equation can be rewritten as

$$\mu (H_r)_{r=\Re^-} - (H_r)_{r=\Re^+} = 0, \tag{56}$$

where $H_r$ is a radial component of the RF magnetic field.

At the same time, as we pointed out above, operator $\hat{L}_\perp$ becomes self-adjoint for homogeneous boundary conditions (continuity of $\tilde{\psi}$ and $\tilde{B}_r$) on a lateral surface. The homogeneous boundary condition for the radial component of $\vec{B}$ in a cylindrical ferrite rod of radius $\Re$ is written as:

$$\mu(H_r)_{r=\Re^-} - (H_r)_{r=\Re^+} = -i\mu_a (H_\theta)_{r=\Re}. \tag{57}$$

The quantity $(H_\theta)_{r=\Re}$ means the azimuth magnetic field on the border circle. For magnetostatic solutions it becomes clear [21] that because of the boundary condition (57) membrane functions cannot be considered as single-valued functions. This fact raises a question about validity of the energy orthogonality relation for MS-wave modes. For a system for which a total Hamiltonian is conserved, there should be single valuedness for egenfunctions. Thus, we are arguing that the internal disk region (where diagonal component of the permeability tensor $\mu < 0$) must have a domain wall at the edge. One can use a notion of a domain wall as a model of the edge. As we will show, in order to cancel the "edge anomaly", the boundary excitation must be described by chiral states.

Following a standard way of solving boundary problems in mathematical physics [29, 30], one can consider two joint boundary problems: the main boundary problem and the conjugate boundary problem. The problems are described by differential equations which are similar to Eq. (10). The main problem is expressed by a differential equation:

$$\left(\hat{L}_\perp - i\beta \hat{R}\right)\tilde{V} = 0. \tag{58}$$

The conjugate problem is expressed by an equation:

$$\left(\hat{L}_\perp^\circ - i\beta^\circ \hat{R}\right)\tilde{V}^\circ = 0. \tag{59}$$

From a formal point of view, it is supposed initially that these are different equations: there are different differential operators, different eigenfunctions and different eigenvalues. A form of differential operator $\hat{L}_\perp^\circ$ one gets from integration by parts:

$$\int_S (\hat{L}_\perp \tilde{V})(\tilde{V}^\circ)^* dS = \int_S \tilde{V}(\hat{L}_\perp^\circ \tilde{V}^\circ)^* dS + \oint_C P(\tilde{V},\tilde{V}^\circ) dC, \tag{60}$$

where $C$ is a contour surrounding a cylindrical ferrite core and $P(\tilde{V},\tilde{V}^\circ)$ is a bilinear form.

Operator $\hat{L}_\perp$ is a self-conjugate operator when permeability tensor $\tilde{\mu}$ is a Hermitian tensor and when a contour integral in the right-hand side of Eq. (60) is equal to zero. The last condition means that for an open ferrite structure [a core ferrite region ($F$) is surrounded by a dielectric region ($D$)] the homogeneous boundary conditions for functions $\tilde{V}$ and $\tilde{V}^\circ$ should give

$$\oint_C P(\tilde{V},\tilde{V}^\circ) dC \equiv \oint_C [P^{(F)}(\tilde{V},\tilde{V}^\circ) + P^{(D)}(\tilde{V},\tilde{V}^\circ)] dC = 0. \tag{61}$$

Since in a ferrite region $\tilde{B}_r = \mu \frac{\partial \tilde{\varphi}}{\partial r} + i\mu_a \frac{\partial \tilde{\varphi}}{\partial \theta}$ and in a dielectric $\tilde{B}_r = \frac{\partial \tilde{\varphi}}{\partial r}$, one has [21]

$$\oint_C P(\tilde{V},\tilde{V}^\circ) dC = \oint_C \left[\mu \left(\frac{\partial \tilde{\varphi}}{\partial r}\right)_{r=\Re^-} - \left(\frac{\partial \tilde{\varphi}}{\partial r}\right)_{r=\Re^+}\right] (\tilde{\varphi}^\circ)^*_{r=\Re} - (\tilde{\varphi})_{r=\Re} \left[\mu \left(\frac{\partial \tilde{\varphi}^\circ}{\partial r}\right)_{r=\Re^-} - \left(\frac{\partial \tilde{\varphi}^\circ}{\partial r}\right)_{r=\Re^+}\right]^* dC +$$

$$\oint_C \left[\left(i\mu_a \frac{\partial \tilde{\varphi}}{\partial \theta}\right)(\tilde{\varphi}^\circ)^* - (\tilde{\varphi})\left(\left(i\mu_a \frac{\partial \tilde{\varphi}}{\partial \theta}\right)^\circ\right)^*\right]_{r=\Re} dC$$

$$\tag{62}$$

The dimensionless membrane function $\tilde{\varphi}$ is considered as a function satisfying boundary conditions (57). Evidently, this function changes a sign when $\theta$ is rotated by $2\pi$. So there are two solutions: $\tilde{\varphi}_+$ and $\tilde{\varphi}_-$. We introduce now the dimensionless membrane function $\tilde{\eta}$ which satisfies boundary condition (56). Since $\hat{G}_\perp$ and $\hat{L}_\perp$ are linear operators, we can write the following relation:

$$\tilde{\eta}(\rho,\alpha) = \begin{cases} \gamma_- \tilde{\varphi}_+, \\ \gamma_+ \tilde{\varphi}_- \end{cases} \tag{63}$$

where

$$\gamma_{\mp} \equiv a_{\mp} e^{-iq_{\mp}\theta}. \tag{64}$$

is an auxiliary function. To preserve the single-valued nature of the membrane function, function $\gamma_{\mp}$ must change its sign when $\theta$ is rotated by $2\pi$ so that $e^{-iq_{\mp}2\pi} = -1$. That is

$$q_{\mp} = \mp l\frac{1}{2}, \tag{65}$$

where $l = 1, 3, 5, ...$ We rewrite Eq. (63) as follows:

$$\tilde{\varphi}_{\pm} = \frac{1}{\gamma_{\mp}}\tilde{\eta} = \delta_{\pm}\tilde{\eta}, \tag{66}$$

where

$$\delta_{\pm} = \frac{1}{a_{\mp}} e^{-iq_{\pm}\theta} \equiv f_{\pm} e^{-iq_{\pm}\theta}. \tag{67}$$

The quantity $q_{\pm}$ is equal to $\pm l\frac{1}{2}$. In the above relations, evidently, $a_{+} = -a_{-}$ and $f_{+} = -f_{-}$. To have proper normalization we will take $|a_{\mp}| = |f_{\pm}| = 1$.

Functions of the conjugate problem have the forms similar to those written for functions of the main problem:

$$\tilde{\varphi}_{\pm}^{\circ} = \frac{1}{\gamma_{\mp}^{\circ}}\tilde{\eta}^{\circ} = \delta_{\pm}^{\circ}\tilde{\eta}^{\circ}, \tag{68}$$

where

$$\delta_{\pm}^{\circ} \equiv f_{\pm}^{\circ} e^{-iq_{\pm}^{\circ}\theta}, \tag{69}$$

$f_{+}^{\circ} = -f_{-}^{\circ}$, $|f_{\pm}^{\circ}| = 1$, and $q_{\pm}^{\circ}$ are half-integer numbers.

The edge states are defined by four edge functions: $\delta_{+}, \delta_{-}, \delta_{+}^{\circ}$, and $\delta_{-}^{\circ}$. Based on a simple analysis (see [21]) one can reduce Eq. (62) to

$$\oint_C P_{\pm}(\tilde{V}, \tilde{V}^{\circ}) dC = \Re \int_0^{2\pi} \left[ \left( i\mu_a \frac{\partial \delta_{\pm}}{\partial \theta} \right)(\delta_{\pm}^{\circ})^* - (\delta_{\pm})\left( \left( i\mu_a \frac{\partial \delta_{\pm}}{\partial \theta} \right)^{\circ} \right)^* \right]_{r=\Re} d\theta. \tag{70}$$

Here, putting signs for a bilinear form $P(\tilde{V}, \tilde{V}^{\circ})$, we took into account the fact that there is the possibility for two [or positive (+), or negative (–)] solutions on the boundary.

A sign of a full chiral rotation, $q_{+}\theta = \pi$ or $q_{-}\theta = -\pi$, should be correlated with a sign of the parameter $i\mu_a$. This becomes evident from the fact that a sign of $i\mu_a$ is related to a precession direction of a magnetic moment $\vec{m}$. In a ferromagnetic resonance, the bias field sets up a preferential precession direction. For a normally magnetized ferrite disk with the bias field directed along $z$ axis, a transverse component of a magnetic moment $\vec{m}$ precesses counterclockwise about the field. Because of the preferential precession direction for a given direction of a bias field, the following condition should be used in our analysis:

$$(i\mu_a)^\circ = i\mu_a. \tag{71}$$

So one obtains

$$\oint_C P_\pm(\widetilde{V},\widetilde{V}^\circ)\, dC = \Re \int_0^{2\pi} \left[\left(i\mu_a \frac{\partial \delta_\pm}{\partial \theta}\right)(\delta_\pm^\circ)^* - (\delta_\pm)\left(i\mu_a \frac{\partial \delta_\pm^\circ}{\partial \theta}\right)^*\right]_{r=\Re} d\theta =$$
$$\Re \mu_a (q_\pm - q_\pm^\circ) \int_0^{2\pi} [\delta_\pm (\delta_\pm^\circ)^*]_{r=\Re}\, d\theta. \tag{72}$$

Fig. 3 illustrates two cases of directions of a chiral rotation in a correlation with directions of the RF magnetization evolution in a ferrite disk. We will call these cases as the (+) resonance (when a direction of a chiral rotation coincides with the precession magnetization direction) and the (–) resonance (when a direction of a chiral rotation is opposite to the precession magnetization direction). The MS-wave membrane function maps to itself under simultaneous change of a sign of the magnetization vector and shift of the phase of function $\delta$ by $+\pi$ [in a case of the (+) resonance] or by $-\pi$ [in a case of the (–) resonance]. It means that for function $\delta$ both the counterclockwise and clockwise directions of going around make physical sense.

From demand of self-adjointness of operator $\hat{L}_\perp$ one has the orthogonality conditions:

$$(q_+ - q_+^\circ) \int_0^{2\pi} [\delta_+ (\delta_+^\circ)^*]_{r=\Re}\, d\theta = 0 \tag{73a}$$

and

$$(q_- - q_-^\circ) \int_0^{2\pi} [\delta_- (\delta_-^\circ)^*]_{r=\Re}\, d\theta = 0. \tag{73b}$$

So there are the normalization relations for the edge functions:

$$\int_0^{2\pi} [\delta_+ (\delta_+^\circ)^*]_{r=\Re}\, d\theta \equiv (N_\theta)_+ \tag{74a}$$

and

$$\int_0^{2\pi} [\delta_- (\delta_-^\circ)^*]_{r=\Re}\, d\theta \equiv (N_\theta)_-, \tag{74b}$$

where $(N_\theta)_+$ and $(N_\theta)_-$ are real quantities. For a given direction of a bias magnetic field, the waves described by functions $\delta_+$ and $\delta_+^\circ$ propagate only in one direction along the edge. Also the waves described by functions $\delta_-$ and $\delta_-^\circ$ propagate in one direction (opposite to the former case) along the edge.

It is necessary to point out that formally one can suppose that together with Eq. (71) the following relation takes place: $(i\mu_a)^\circ = -i\mu_a$. So such orthogonality conditions as $(q_\pm + q_\mp^\circ) \int_0^{2\pi} [\delta_\pm (\delta_\mp^\circ)^*]_{r=\Re} d\theta = 0$ may occur. These conditions, however, are beyond a physical meaning. Based on such conditions one cannot restore singlevaluedness (and, therefore, Hermicity) of the spectral problem. Any observation of the vector potential terms, the gauge fields, and the fluxes (considered below) is excluded in this case.

## 4. The Aharonov-Bohm-like effect for dipolar-mode magnetization motion in a ferrite disk

The vector potential is considered to be nonobservable in Maxwellian electromagnetism. At the same time, the vector potential can be observable in the Aharonov-Bohm [31] or Aharonov-Casher [32] effects, but only via its line integral, not pointwise. The above analysis shows that a line integral $\oint_C \left( i \frac{\partial \delta_\pm}{\partial \theta} \right) (\delta_\pm^\circ)^* dC = \Re \int_0^{2\pi} \left[ \left( i \frac{\partial \delta_\pm}{\partial \theta} \right) (\delta_\pm^\circ)^* \right]_{r=\Re} d\theta$ is an observable quantity. It follows from the fact that because of such a quantity one can restore singlevaluedness (and, therefore, Hermicity) of the spectral problem. We can represent this observable quantity as a linear integral of a certain vector potential.

Let us consider an integral $\int_0^{2\pi} \int_S (i\vec{\nabla}_\theta \tilde{\varphi}_p)(\tilde{\varphi}_p^\circ)^* dS\, d\theta$, which corresponds to certain mode $p$. This integral, being a real quantity because of the normalization conditions, can be represented as (we omitted now index $p$):

$$\int_0^{2\pi} \int_S (i\vec{\nabla}_\theta \tilde{\varphi})(\tilde{\varphi}^\circ)^* dS\, d\theta = \int_0^{2\pi} \int_S [\delta_\pm (\delta_\pm^\circ)^*][(i\vec{\nabla}_\theta \tilde{\eta})(\tilde{\eta}^\circ)^*] dS\, d\theta +$$
$$\int_0^{2\pi} \int_S [\tilde{\eta}(\tilde{\eta}^\circ)^*][(i\vec{\nabla}_\theta \delta_\pm)(\delta_\pm^\circ)^*] dS\, d\theta. \tag{75}$$

The first integral in the right-hand side (RHS) of Eq. (75) is the dynamical phase factor for an oscillating mode, but the second integral in the RHS of this equation is the geometrical phase factor. The geometrical phase factor is not single-valued under continuation around a circuit and can be correlated with the vector potential [33]:

$$i\Re \int_0^{2\pi} [(\vec{\nabla}_\theta \delta_\pm)(\delta_\pm^\circ)^*]_{r=\Re} d\theta \equiv \oint_C \left( \vec{A}_\theta^m \right)_\pm \cdot d\vec{C} = 2\pi q_\pm. \tag{76}$$

The Berry's phase is generated from the broken dynamical symmetry.

The confinement effect for magnetic-dipolar oscillations requires proper phase relationships to guarantee single-valuedness of the wave functions. To compensate for sign ambiguities and thus to make wave functions single valued we added a vector-potential-type term to the MS-potential Hamiltonian. A circulation of vector $\vec{A}_\theta^m$ should enclose a certain flux. The corresponding flux of pseudo-electric field $\vec{\in}$ (the gauge field) through a circle of radius $\Re$ is obtained as:

$$\int_S (\vec{\in})_\pm \cdot d\vec{S} = \oint_C (\vec{A}_\theta^m)_\pm \cdot d\vec{C} = (\Xi^e)_\pm, \tag{77}$$

where $(\Xi^e)_\pm$ is the flux of pseudo-electric field. There should be the positive and negative fluxes. These different-sign fluxes should be inequivalent to avoid the cancellation. This is the case of chirality: the chirality breaks time-reversal symmetry and the positive and negative vortices are not equivalent. Our analysis shows that only the phase factor $\exp[i(\Xi^e)_\pm]$ and not the phase $(\Xi^e)_\pm$ is physically meaningful. An examination of the AB experiment indicates the same fact [34]. Superscript "$m$" in the vector-potential term means that there is the physical quantity associated with the magnetization motion, or the magnetic-current vector potential (see Appendix). In our problem, the vector potential $A_\theta^m$ is due to a surface magnetic current. Because of an irrotational flow on a border contour $C$, one has $\vec{\nabla} \times \vec{A}_\theta^m = 0$. It means that an electric field on contour $C$ is equal to zero. But the vector potential can be observable via its line integral. A magnetic moment moving along contour $C$ in the gauge field feels no force and undergoes the Aharonov-Bohm-type interference effect.

## 5. Edge magnetic currents, anapole moments, and the magnetoelectric energy splitting in a ferrite disk particle

The most basic implication of the existence of a phase factor in $\tilde{\varphi}$ is operative in the case on the border ring region. The single-valuedness of mode membrane functions requires that this phase factor returns to itself modulo $2\pi$ on going twice around the circuit. The solutions for functions $\tilde{\varphi}$ depend on both a sign of functions $\delta$ and a sign of $\mu_a$. So one can distinguish formally *four solutions* for functions $\tilde{\varphi}$. If the gauge field is viewed as a part of the internal dynamics of the system then the four solutions should correspond to four states in a single-Hilbert space.

The above four solutions arise from four functions: $\delta_\pm$ and $\delta_\pm^\circ$. Because of the preferential precession direction for a given direction of a bias field, condition (71) should be taken into account. Therefore, for a given direction of a bias field there are two edge modes: one corresponding to the $(+)$ resonance (when $\delta = \delta_+ = \delta_+^\circ$) and another corresponding to the $(-)$ resonance (when $\delta = \delta_- = \delta_-^\circ$). The topological effects are manifested through the generation of relative phases which accumulate on the boundary wave function $\delta$.

In the spectral problem under consideration, the boundary relations are the edge states which are separate from the cross-sectional states. As we discussed above, the difference between the boundary conditions for self-adjointess of operator $\hat{G}_\perp$ and self-adjointess of operator $\hat{L}_\perp$ is due to a non-zero border term in the right-hand side of Eq. (57):

$$\text{border term} \equiv -i\mu_a (H_\theta)_{r=\Re}. \tag{78}$$

It will be shown below that this border term can be observable via its circulation integral.

Taking into account Eqs. (2) and (66), we can represent an annual magnetic field $H_\theta$ in this term as follows

$$\left(\left(\vec{H}_\theta(z)\right)_\pm\right)_{r=\Re} = -\xi(z)\vec{\nabla}_\theta \tilde{\varphi}_\pm = -\xi(z)\left(\tilde{\eta}\vec{\nabla}_\theta \delta_\pm + \delta_\pm \vec{\nabla}_\theta \tilde{\eta}\right), \tag{79}$$

where function $\xi(z)$ characterizes $z$-distribution of the MS potential in a ferrite disk [22]. It is evident that circulation of gradient

$$\vec{\nabla}_\theta \delta_\pm = \frac{1}{\Re}\left.\frac{\partial \delta_\pm}{\partial \theta}\right|_{r=\Re} \vec{e}_\theta = -i\frac{q_\pm f_\pm}{\Re} e^{-iq_\pm \theta}\vec{e}_\theta \tag{80}$$

along contour $C$ gives a nonzero quantity when $q_\pm$ is a number divisible by $\frac{1}{2}$. On the contrary, circulation of gradient $\vec{\nabla}_\theta \tilde{\eta}$ along contour $C$ is equal to zero. We consider the quantity $\nabla_\theta \delta_\pm$ as the velocity of an irrotational "border" flow:

$$\left(\vec{V}_\theta\right)_\pm \equiv \vec{\nabla}_\theta \delta_\pm. \tag{81}$$

In such a sense, functions $\delta_\pm$ are the velocity potentials [35].

The velocity field arises from the Berry phase of the spin field. Fig. 3 (a) corresponds to the $(+)$ resonance with velocity $\left(\vec{V}_\theta\right)_+$ of a chiral rotation and Fig. 3 (b) corresponds to the $(-)$ resonance with velocity $\left(\vec{V}_\theta\right)_-$ of a chiral rotation. The non-zero circulation of the velocity $\left(\vec{V}_\theta\right)_\pm$ along contour $C$ – the border ring region of a ferrite disk – may have a physical meaning of vorticity [35]. Two cases of chiral rotations described by vectors $\left(\vec{V}_\theta\right)_\pm$ can be characterized by vectors of angular moments directed normally to a disk. For a given cross-sectional state (a cross-sectional mode described by membrane function $\tilde{\eta}$), we define the strength of a vortex of a *whole disk* as

$$s_\pm^e \equiv \tilde{\eta}\int_0^d \xi(z)dz \oint_C \left(\vec{V}_\theta\right)_\pm \cdot d\vec{C} = \Re\tilde{\eta}\int_0^d \xi(z)dz \int_0^{2\pi} \nabla_\theta \delta_\pm \, d\theta = -2f_\pm \tilde{\eta}\int_0^d \xi(z)dz, \tag{83}$$

where $d$ is a disk thickness. A physical meaning of a superscript "$e$" in a designation of $\vec{s}_\pm^e$ will be explained below.

We define now an angular moment $\vec{a}_\pm^e$:

$$a_\pm^e \equiv \int_0^d \oint_C (\text{border term})\, \vec{e}_\theta \cdot d\vec{C}\, dz = i\mu_a s_\pm^e. \tag{84}$$

This angular moment can be formally represented as a result of a circulation of a quantity, which we call a density of an effective boundary magnetic current $\vec{i}^{\,m}$:

$$a_\pm^e = 4\pi \int_0^d \oint_C \vec{i}_\pm^{\,m} \cdot d\vec{C}\, dz, \tag{85}$$

where

$$\vec{i}_{\pm}^{\,m} \equiv \rho^m \left(\vec{V}_\theta\right)_\pm \tag{86}$$

and

$$\rho^m \equiv i\frac{\mu_a}{4\pi}\xi\tilde{\eta}. \tag{87}$$

Relation (85) can be considered as generalization of the Ampere hypothesis.

The current $\vec{i}^{\,m}$ is a persistent magnetic current in an Aharonov-Bohm-like geometry. On an edge ring region, we have a magnetization motion pierced by an electric flux $\left(\Xi^e\right)_\pm$. In contrast to the spin-transport analysis of persistent spin current in a mesoscopic ferromagnetic Heisenberg ring in an uniform magnetic field [36], in our continuous-medium model a character of the magnetization motion becomes apparent via the gyration parameter $\mu_a$ in the boundary term for the spectral problem [see Eq. (62)]. There is magnetization motion through a non-simply-connected region. On the edge region, the chiral symmetry of the magnetization precession is broken to form a flux-closure structure. The edge magnetic currents can be observable only via its circulation integrals, not pointwise. This results in the moments oriented along a disk normal. It was shown experimentally [37] that such a moment has a response in an external RF electric field and so can be called as an electric moment. This clarifies a physical meaning of a superscript "e" in designations of $s_\pm^e$ and $a_\pm^e$.

Let us analyze mutual correlations of vectors $\vec{a}^e$ and $\vec{s}^e$ for a ferrite disk particle. These vectors are directed along $+z$ or $-z$ axis. An analysis of the time-reversal and space-inversion properties of these configurations will clarify the symmetry properties of the edge states. It is evident that for a given type of a resonance [the (+) or (–) resonance], vectors $\vec{a}^e$ and $\vec{s}^e$ can be mutually parallel or anti parallel. These two cases are equally possible. The case of parallel vectors $\vec{a}^e$ and $\vec{s}^e$ we will conventionally call the "antiparticle" configuration, while the case of anti parallel vectors $\vec{a}^e$ and $\vec{s}^e$ – the "particle" configuration.

Let a bias magnetic field be directed along $+z$ axis. For the (+) resonance we have vector $\vec{s}_+^e$ directed along $+z$ axis, while for the (–) resonance we have vector $\vec{s}_-^e$ directed along $-z$ axis. Suppose that in both cases we have the "antiparticle" configurations Fig. 4. Because of the time-reversal process, the (+) resonance cannot be transformed to the (–) resonance and vice versa. This results in to a fact that direction of vector $\vec{a}^e$ is invariant under the time reversal. So for the time-reversal process shown in Fig. 4, one has transformations of the "antiparticle" configurations to the "particle" configurations.

In a ferrite disk particle, the vector $\vec{a}^e$ can be characterized as an electric moment with the anapole-moment properties. If we classically picture an element of the magnetic current $i^m$ as a small electric-current loop, the combination of all elements along a disk border contour $C$ can be viewed as a toroidal electric-current winding [38, 39]. One can trace a clear analogy between properties of our structure and the electromagnetic properties of a toroidal solenoid. Following Zel'dovich, such a system has a parity-odd toroidal (or anapole) moment [38, 39]. The toroid dipole (being originated from a toroidally-wound solenoid shrunken to a point) singles out a direction in space exactly as does the usual electric dipole. Because of the parity violation, vectors $\vec{a}^e$ become directed in an opposite way for the (+) and (–) resonances. Fig. 5 shows transformations of the "antiparticle" configurations to the "particle" configurations due to the space-inversion process for the (+) and (–) resonances. For a certain resonance [the (+) resonance or the (–) resonance], an analysis of the *PT* symmetry of the pictures in the Figs. 4 and 5 shows invariance under such a combined operation. This gives another proof of the Hermicity of the spectral problem [40]. The *PT* invariance allows excitation of the edge persistent magnetic current in the absence of the external electric flux.

It is evident that a complex conjugate "particle" configuration is an "antiparticle" configuration and vice versa. This fact is related to the existence of only a single edge-mode excitation for each wavenumber $q$. In other words, the "particle" configurations are their own "antiparticle" configurations. This resembles the well known properties of the Majorana fermions [26]. The edge wave propagates only in one direction along the cylindrical surface of a disk. So we have a "chiral Majorana fermion field" [41, 42] on the lateral wall of a disk.

The demand for single valuedness in the above treatment has been made for the total wavefunction describing a closed system, i.e. a system for which the total Hamiltonian is conserved. However, for scattering of electromagnetic fields by a thin-ferrite-disk particle we have a Hamiltonian which is a function of time. For such an open system, there cannot be the requirement of single valuedness. If the gauge field (a flux) is viewed as external, it should be fixed as a part of the definition of the problem and there will be different energy states for the possible choices of boundary conditions. What picture should appear when one considers scattering of electromagnetic fields by a thin-ferrite-disk particle? For a certain energy level of mode $p$ corresponding to a closed system [see Eq. (25)], there should be positive and negative splits of energy (with respect to an energy level of mode $p$) for the "particle" and "antiparticle" configurations. The solutions with the positive and negative energy splits are the creation and annihilation operators for the same $p$-mode energy level. Functions $\tilde{\varphi}$ are vortex complex fields. In such a sense, $\tilde{\varphi}$ and $\tilde{\varphi}^*$ can be viewed as vortex quantum field operators, which destroy and create vortices. We can see that the creation and annihilation operators are operators of the same $p$-mode light magnon. From the above analysis it follows that different orientations of an electric moment $\vec{a}^e$ (parallel or antiparallel with respect to $\vec{H}_0$) correspond to different energy levels. In an open system, the energy splitting between two cases: $\vec{a}^e \cdot \vec{H}_0 > 0$ and $\vec{a}^e \cdot \vec{H}_0 < 0$ we can characterize as the *magnetoelectric* energy splitting.

## 6. Discussion and conclusion

In ferrite disks, the microwave vortices can appear in different kinds of physical phenomena depending on space scales of the wave processes. From one side, one can observe the exchange-interaction vortices of magnetization [8, 15]. From the other side, there are the electromagnetic-wave (Poynting vector) vortices [43]. Space scales of the MS-wave processes are much less than wavelengths of electromagnetic waves and much bigger than wavelengths of exchange waves. Physically, the magnetic-dipolar-mode oscillations are neither electromagnetic-wave nor exchange-wave oscillations [21, 24, 28]. In this paper we showed unique features of the MS-mode vortices.

For magnetic-dipolar modes in a normally magnetized thin ferrite disk, the border-region interaction can be interpreted in terms of a chiral contribution to the precessing-electron magnetization. In order to cancel non-singlevaluedness of the magnetic-dipolar modes, the boundary excitations must be described by chiral states. The chiral states appear as rings of vortices moving around the apparently undisturbed membrane-function discrete-energy states.

The chiral edge states constitute a subset of the spectrum. A feature of the Dirac-like nature of the chiral edge spectrum is that there are two kinds of modes with the "particle" and "antiparticle" configurations. This can be described by saying that there are half-integer quantum numbers of the vortices ("charges") and that "particle" and "antiparticle" configurations carry opposite values of these quantum numbers. This is the case of dynamical symmetry breaking. The chirality breaks time reversal symmetry, and the positive and negative vortices are not equivalent. Appropriate quasiparticles are just effective fields describing the collective excitations on the edge and are considered in terms of which a complicated strongly interacting system appears weakly interacting.

The gauge field (fluxes) can be viewed either as external or as a part of the internal dynamics of the system. The demand for single valuedness in our treatment has been made for the total MS-potential function describing a closed system, i.e. a stationary-state system for which the total Hamiltonian is conserved. For an open system – the system interacting with an external electromagnetic field – we do not have the requirement of singlevaluedness. Such edge states existing in the thin-walled cylinder and having a certain handedness, can be excited only by increasing (decreasing) the angular moment of a whole ferrite disk [37].

**Appendix: Electric-current and magnetic-current vector potentials**

In a case of the "true" AB effect we have the electric-current vector potential resulting, in fact, from the electric-charge motion. In a case of the MDM AB-like effect we have the magnetic-current vector potential resulting from the magnetization motion. The "true" AB effect comes from the gauge invariant coupling between the electric current and the electric-current vector potential. For that reason this effect can be observed even in the absence of the magnetic field. The MDM AB-type interference comes from the gauge invariant coupling between the "magnetic" current and the magnetic-current vector potential.

In the electromagnetic field theory, the motion equations of a charge in the electromagnetic field give correlations between the $\vec{E}$ and $\vec{B}$ fields and the vector $\vec{A}$ and scalar $\varphi$ potentials:

$$\vec{E} = -\frac{1}{c}\frac{\partial \vec{A}}{\partial t} - \nabla \varphi; \qquad \vec{B} = \nabla \times \vec{A}. \qquad (A1)$$

The same relations one immediately obtains from the Maxwell equations. For the electric current density $\vec{j}^e$, as a source of the electromagnetic field, one has the wave equation for vector potential $\vec{A}$ [9]:

$$\nabla^2 \vec{A} - \frac{1}{c^2}\frac{\partial^2 \vec{A}}{\partial t^2} = -\frac{4\pi}{c}\vec{j}^e. \qquad (A2)$$

In spite the fact that no magnetic charges and no motion equations for magnetic charges are known in nature, because of the electromagnetic duality one can formally introduce magnetic currents in Maxwell equations. This formal procedure allows solving numerous electrodynamics problems [44, 45]. To distinguish an electrodynamic vector potential caused by an electric current from an electrodynamic vector potential caused by a magnetic current, we rewrite Eq. (A2) as

$$\nabla^2 \vec{A}^e - \frac{1}{c^2}\frac{\partial^2 \vec{A}^e}{\partial t^2} = -\frac{4\pi}{c}\vec{j}^e, \qquad (A3)$$

where $\vec{A}^e$ means the electric-current vector potential. Together with this wave equation one has from Maxwell equations another wave equation [44, 46]:

$$\nabla^2 \vec{A}^m - \frac{1}{c^2}\frac{\partial^2 \vec{A}^m}{\partial t^2} = -\frac{4\pi}{c}\vec{j}^m, \tag{A4}$$

where $\vec{A}^m$ means the magnetic-current vector potential.

**References**

[1] A. Hubert and R. Schäfer, *Magnetic Domains* (Springer, Berlin, 1998).
[2] T. Shinjo *et al.*, Science **289**, 930 (2000).
[3] K.Yu. Guslienko *et al*, Phys. Rev. B **65**, 024414 (2001).
[4] J. M. Luttinger and L. Tisza, Phys. Rev. **70**, 954 (1946).
[5] H. Puszkarski, M. Krawczyk, and J.-C. S. Levy, Phys. Rev. B **71**, 014421 (2005).
[6] P.H. Bryant *et al*, Phys. Rev. B **47**, 11255 (1993).
[7] K.Yu. Guslienko *et al*, Phys. Rev. B **66**, 132402 (2002).
[8] C.E. Zaspel *et al*, Phys. Rev. B **72**, 024427 (2005).
[9] J.D. Jackson, *Classical Electrodynamics*, 2$^{nd}$ ed., Ch. 5 (Wiley, New York, 1975).
[10] P.M. Morse and H. Feshbach, *Methods of Theoretical Physics*, Ch. 8 (McGraw-Hill, New York, 1953).
[11] M. Grimsditch *et al*, Phys. Rev. B **70**, 054409 (2004).
[12] A.G. Gurevich and G.A. Melkov, *Magnetic Oscillations and Waves* (CRC Press, New York, 1996).
[13] This is slightly akin physics of the electromagnetic wave process in an ordinary transmission-line system. In this case the electromagnetic-wave propagation causes space-time fluctuations of a conductivity current in metal parts and an electric displacement current in dielectrics of a transmission line, but there are no spectral problems for the electric-charge-density waves. So the boundary conditions are imposed on the electromagnetic field components and not on the RF currents.
[14] D.D. Osheroff and M.C. Cross, Phys. Rev. Lett. **59**, 94 (1987).
[15] J. Miltat and A. Thiaville, Science **298**, 555 (2002).
[16] D.D. Stancil, *Theory of Magnetostatic Waves* (Springer-Verlag, New York, 1992).
[17] A.K. Ganguly and D.C. Webb, IEEE Trans. Microw. Theory Tech. **MTT-23**, 998 (1975).
[18] P.R. Emtage, J. Appl. Phys. **49**, 4475 (1978).
[19] A.D. Boardman *et al*, Phys. Rev. E **71**, 026614 (2005).
[20] E.O. Kamenetskii, J. Magn. Magn. Mater. **302**, 137 (2006).
[21] E.O. Kamenetskii, Phys. Rev. E **73**, 016602 (2006).
[22] E.O. Kamenetskii, Phys. Rev. E **63**, 066612 (2001).
[23] A.I. Akhiezer, V.G. Bar'yakhtar, and S.V. Peletminskii, *Spin Waves* (North-Holland, Amsterdam, 1968).
[24] E.O. Kamenetskii, M. Sigalov, and R. Shavit, J. Phys.: Condens. Matter **17**, 2211 (2005).
[25] L.D. Landau and E.M. Lifshitz, *Quantum Mechanics: Non-Relativistic Theory*, 3$^{rd}$ ed. (Pergamon, Oxford, 1977).
[26] A.S. Davydov, *Quantum Mechanics*, 2$^{nd}$ ed. (Pergamon, Oxford, 1976).
[27] R.I. Joseph and E. Schlömann, J. Appl. Phys. **32**, 1001 (1961).
[28] E.O. Kamenetskii, R. Shavit, and M. Sigalov, Europhys. Lett. **64**, 730 (2003).
[29] S. G. Mikhlin, *Variational Metods in Mathematical Physics* (Mc Millan, New York, 1964).
[30] M.A. Naimark, *Linear Differential Operators* (F. Ung. Pub. Co., New York, 1967).
[31] Y. Aharonov and D. Bohm, Phys. Rev. **115**, 485 (1959).
[32] Y. Aharonov and A. Casher, Phys. Rev. Lett. **53**, 319 (1984).
[33] M.V. Berry, Proc. R. Soc. Lond. A **392**, 45 (1984).

[34] T.T. Wu and C.N. Yang, Phys. Rev. B **12**, 3845 (1975).
[35] T. Shimbori and T. Kobayashi, J. Phys. A: Math. Gen. **33**, 7637 (2000).
[36] F. Schütz, M. Kollar, and P. Kopietz, Phys. Rev. Lett. **91**, 017205 (2003).
[37] E.O. Kamenetskii, A.K. Saha, and I. Awai, Phys. Lett. A. **332**, 303 (2004).
[38] Ia. B. Zel'dovich, J. Exp. Theor. Phys. (U.S.R.R.) **33**, 1531 (1957).
[39] G,N, Afanasiev and V.M. Dubovik, J. Phys. A: Math. Gen. **25**, 4869 (1992).
[40] C. Yuce, A. Kurt, and A. Kucukaslan, e-print quant-ph/0602103.
[41] N.Read and D. Green, Phys. Rev. B **61**, 10267 (2000).
[42] D.A. Ivanov, Phys. Rev. Lett. **86**, 298 (2001).
[43] E.O. Kamenetskii, M. Sigalov, and R. Shavit, Phys. Rev. E **74**, 036620 (2006).
[44] J.A. Stratton, *Electromagnetic Theory* (McGraw-Hill, New York, 1941).
[45] H.A. Bethe, Phys. Rev. **66**, 163 (1944).
[46] G. T. Markov and A. F. Chaplin, *Excitation of Electromagnetic Waves* (Radio I Svyaz, Moscow, 1983) (in Russian).

**Figure captions**

FIG. 1. Dependence of quantity $\frac{U^{eff}}{E}$ on a bias magnetic field for magnetic-dipolar oscillating modes in a ferrite disk

FIG.2. A model illustrating the levels of the "kinetic" and potential energies for magnetic-dipolar oscillating modes in a ferrite disk

FIG. 3. Directions of an edge-function chiral rotation in a correlation with directions of the RF magnetization $\vec{m}$ evolution in a ferrite disk. (a) The (+) resonance; (b) the (−) resonance

FIG. 4. Time reversal properties of configurations of vectors $\vec{a}^e$ and $\vec{s}^e$ for the (+) and (−) resonances in correlation with directions of the RF magnetization $\vec{m}$ evolution in a ferrite disk

FIG. 5. Space inversion properties of configurations of vectors $\vec{a}^e$ and $\vec{s}^e$ for the (+) and (−) resonances in correlation with directions of the RF magnetization $\vec{m}$ evolution in a ferrite disk

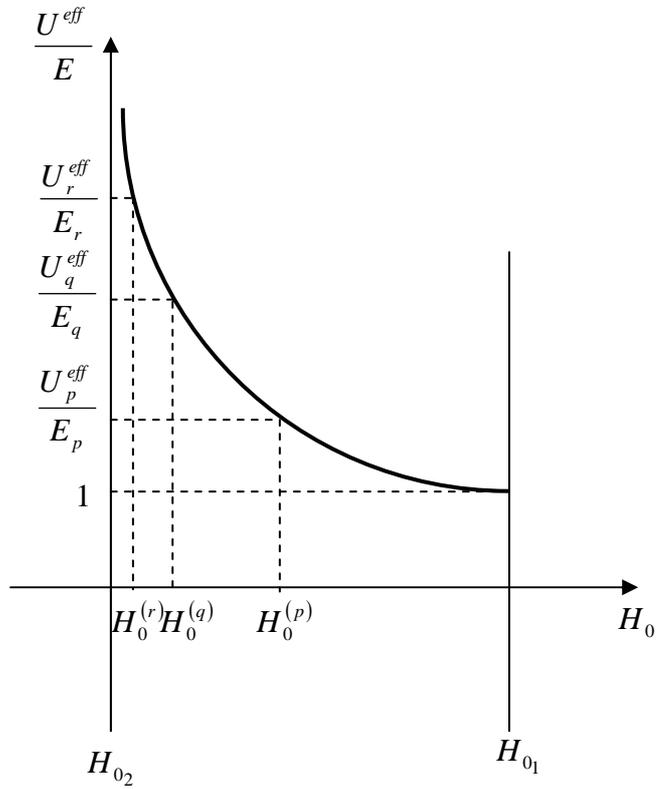

FIG. 1. Dependence of quantity $\dfrac{U^{eff}}{E}$ on a bias magnetic field for magnetic-dipolar oscillating modes in a ferrite disk

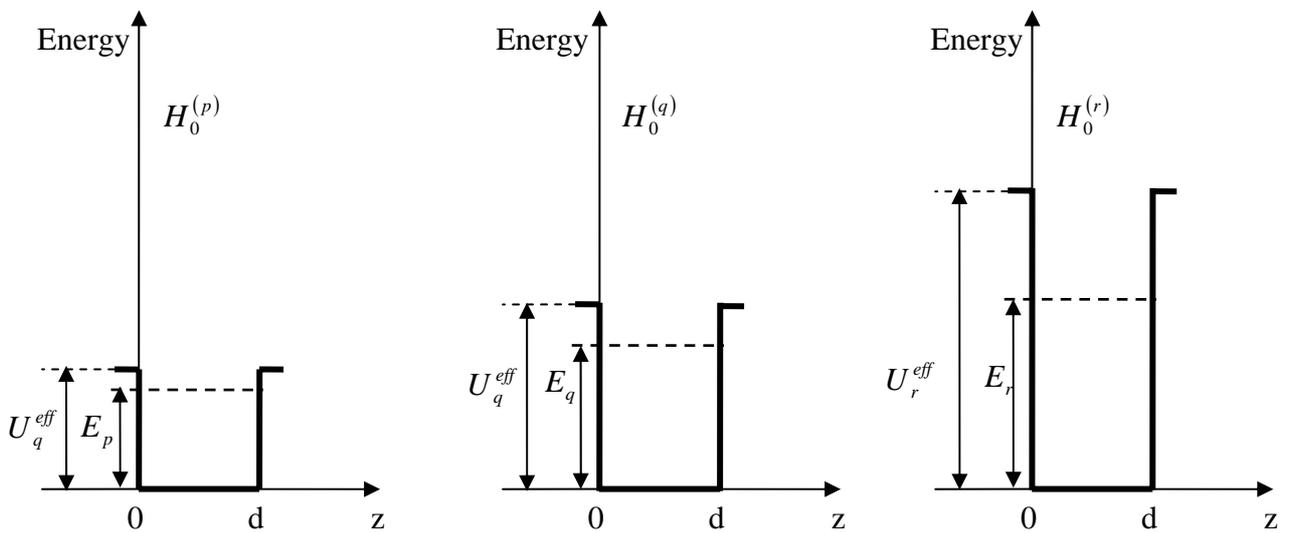

FIG.2. A model illustrating the levels of the "kinetic" and potential energies for magnetic-dipolar oscillating modes in a ferrite disk

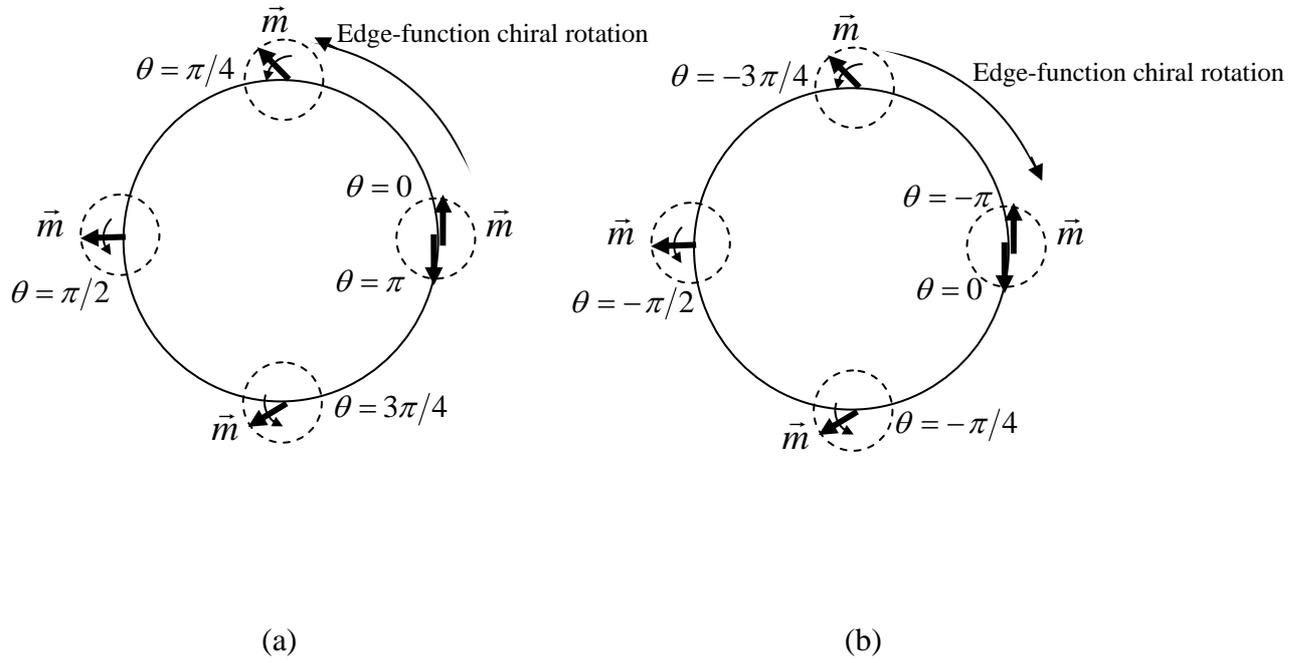

FIG. 3. Directions of an edge-function chiral rotation in a correlation with directions of the RF magnetization $\vec{m}$ evolution in a ferrite disk. (a) The $(+)$ resonance; (b) the $(-)$ resonance

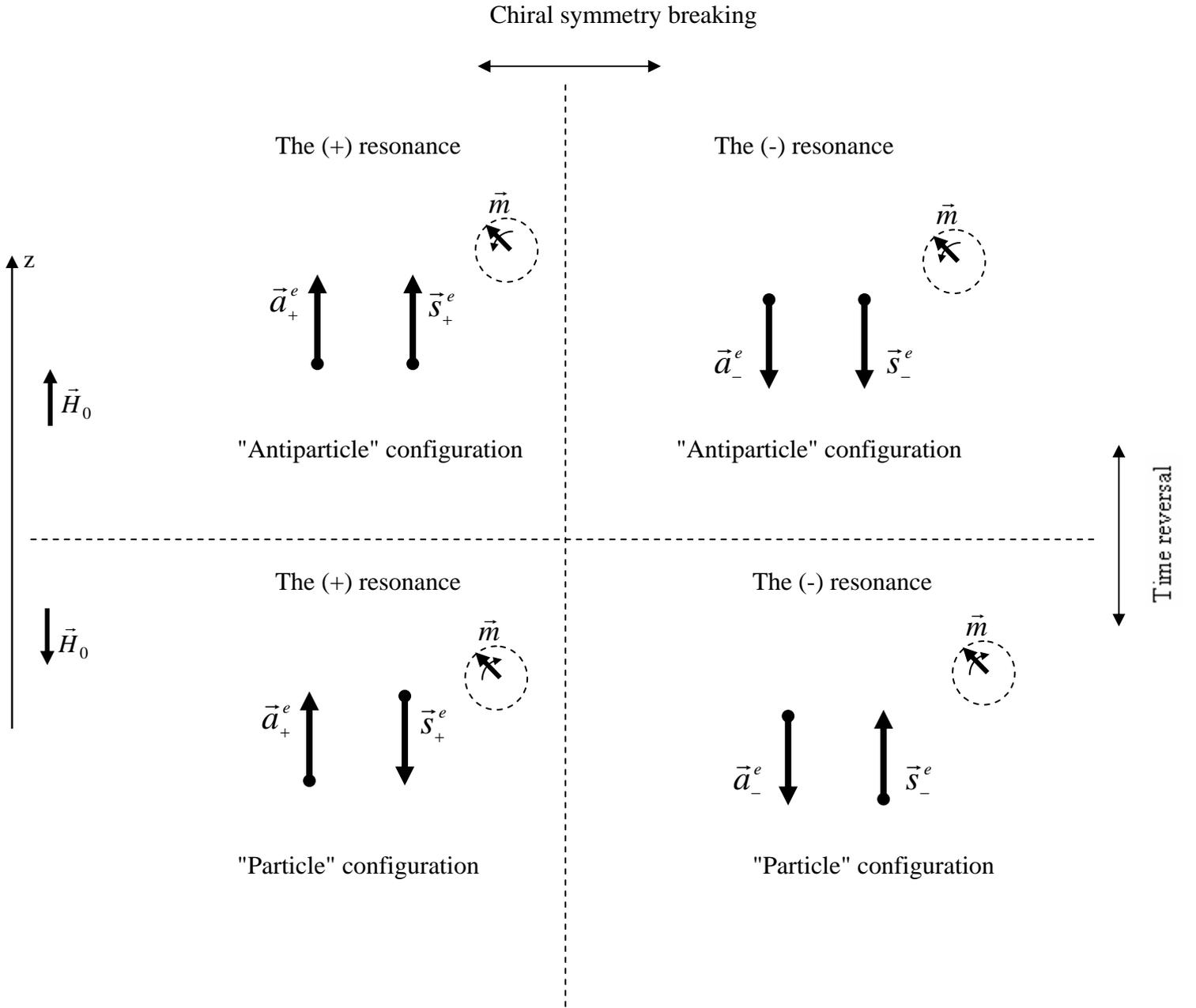

FIG. 4. Time reversal properties of configurations of vectors $\vec{a}^e$ and $\vec{s}^e$ for the $(+)$ and $(-)$ resonances in correlation with directions of the RF magnetization $\vec{m}$ evolution in a ferrite disk

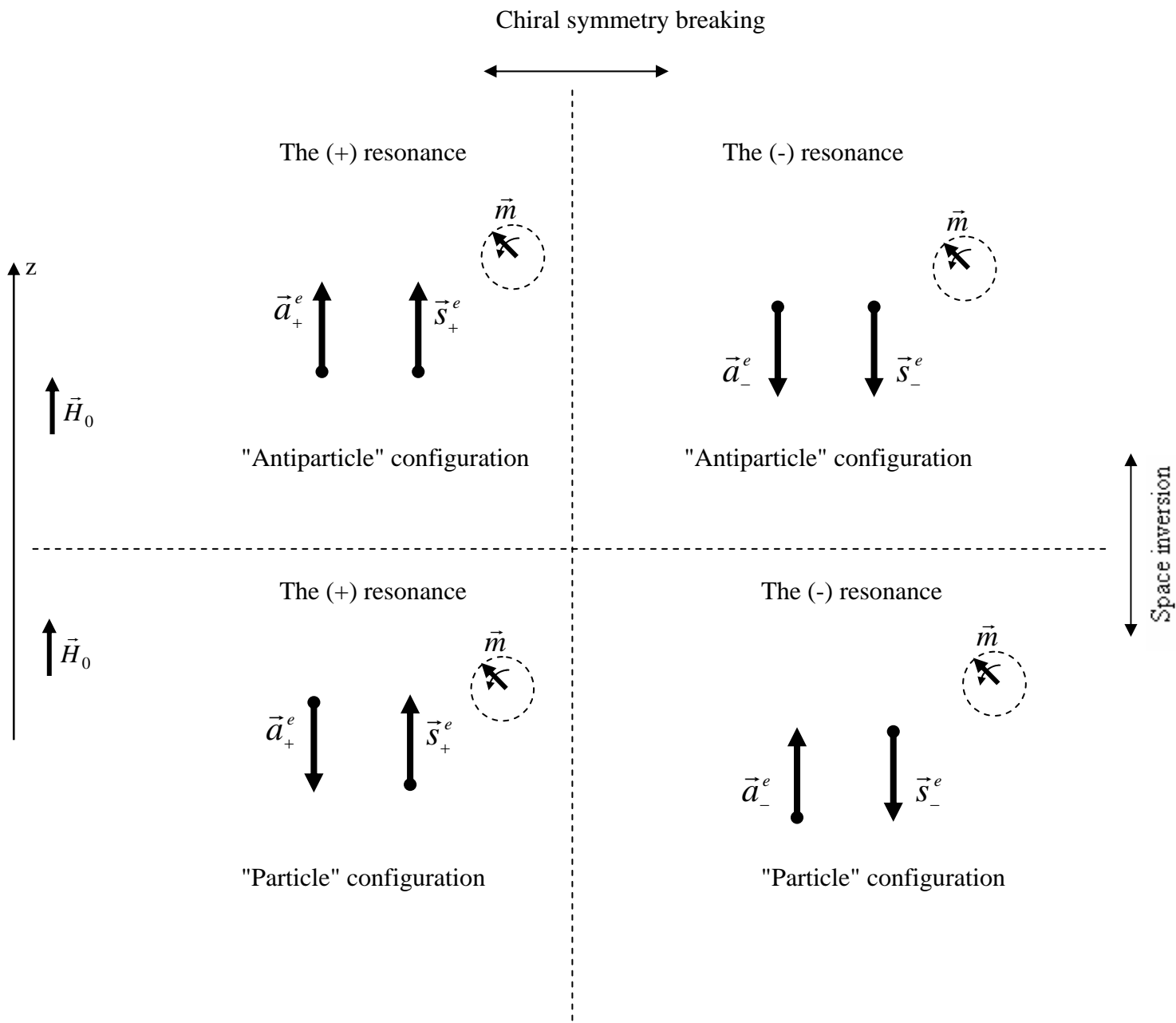

FIG. 5. Space inversion properties of configurations of vectors $\vec{a}^e$ and $\vec{s}^e$ for the (+) and (−) resonances in correlation with directions of the RF magnetization $\vec{m}$ evolution in a ferrite disk